\documentclass[journal,10pt]{IEEEtran}
\usepackage{graphicx}
\usepackage{amsmath}
\usepackage{amssymb}
\usepackage{bm}
\usepackage{algorithm}
\usepackage{algorithmic}
\usepackage{tabularx}
\usepackage{subcaption}

\begin{document}
\title{A Decoupling-based Approach for Signature Estimation of Wideband XL MIMO-FMCW Radars}

\author{$\mathrm{Chandrashekhar~Rai}^{1}$, $\mathrm{Dibbendu~Roy}^{2}$, and $\mathrm{Debarati~Sen}^{3}$ 

\thanks{
    $\mathrm{CS}^1$ is affiliated to Electrical Engineering, Indian Institute of Technology Delhi, India. $\mathrm{DR}^2$ is affiliated to Electrical Engineering, Indian Institute of Technology Indore, India. $\mathrm{DS}^3$ is affiliated to G.S.Sanyal School of Telecommunications, Indian Institute of Technology Kharagpur, India. Email: \emph{csrai.cstaff@iitd.ac.in, droy@iiti.ac.in debarati@gssst.iitkgp.ac.in}.}
\thanks{The conference precursor to this work has been published in the Vehicular Technology Conference (VTC) 2023.}
}

\maketitle

\begin{abstract}
    Modern radars employing wideband signals and extremely large (XL) multiple-input multiple-output (MIMO) arrays can significantly improve range and angular resolution. However, when large bandwidth and array aperture are used simultaneously, the spatial delay across the array becomes comparable to the radar range resolution, leading to the spatial wideband effect (SWE). The SWE introduces several distortions including range migration (range squint), beam squint, and range-angle coupling (RAC), which spread the target response in the range-angle domain and may cause physically separated targets to overlap and mask each other. In this work, we propose a decoupling-based target detection and parameter estimation framework for MIMO frequency modulated continuous wave (FMCW) radar. The proposed method reformulates the joint range-angle estimation problem as a decoupled sequential frequency estimation problem, where the two-dimensional (2D) estimation is carried out through successive one-dimensional (1D) super-resolution estimations. Specifically, we employ orthogonal matching pursuit (OMP) to perform sparse recovery-based range and angle estimation with high resolution. The proposed decoupling strategy is further extended to spatial wideband XL-MIMO FMCW radar systems, enabling reliable detection and separation of targets even when their responses overlap due to severe RAC. Simulation results demonstrate that the proposed approach accurately detects multiple targets and successfully resolves overlapping target responses in the presence of SWE, outperforming conventional Fourier transform and clustering-based methods.
\end{abstract}


\section{Introduction}

Autonomous driving systems rely heavily on high-resolution perception technologies to ensure safe navigation in complex and dynamic environments. Among various sensing modalities, imaging radars have emerged as a critical complement to optical and LiDAR-based sensors, offering robustness under adverse weather, low-visibility, and challenging illumination conditions \cite{hasch2012millimeter,patole2017automotive,waldschmidt2021automotive}. To meet the stringent requirements of fine-grained object localization and scene understanding, next-generation automotive radar systems are goint to be increasingly employing wideband signals and extremely large (XL) multiple-input multiple-output (MIMO) antenna arrays. This combination enables unprecedented range and angular resolution, thereby enhancing the ability to detect closely spaced objects, pedestrians, and road infrastructure \cite{zhang2021mrpt}.

To improve range and angle resolution in MIMO frequency modulated continuous wave (FMCW) radars, extensive research has been conducted in recent years on utilizing high bandwidth and a large number of antenna elements \cite{ding2022tdm,dvorsky2023multistatic,kong2024survey}. However, the simultaneous use of high bandwidth and large-scale arrays introduces the spatial wideband effect (SWE), a phenomenon where the spatial delay across the array becomes comparable to the radar’s range resolution \cite{park2024spatial,rai2025low}. The SWE causes three induced effects in different domains. In the space-range domain, the range bins are significantly migrated across the entire aperture of the extremely large (XL) array aperture, which is known as the range-squint effect. In the angle-frequency domain, the angles observed over a particular frequency sub-band is different than the other frequency sub-band, and hence the angles are squinted through the entire BW, which is known as the beam-squint effect. Morover, in the angle-range domain, the target response is spread in the both angle and range domain jointly, which is known as the range angle coupling (RAC) effect \cite{rabaste2013signal,durr2020range,han2023range,hu2023range}. A similar effect in MIMO orthogonal frequency division multiplexing (OFDM) wireless systems is termed the dual wideband effect \cite{cai2017beamforming,wang2018spatial,wang2019beam,rai2022signature}. The RAC effect (or SWE) depends mainly upon signal bandwidth, array size, and direction of arrival (DoA). For example, the targets at off-boresight angles face more spreading and squinting effects. 

Morover, in narrowband MIMO-FMCW radars, it can be shown that the range-angle estimation can be formulated as a joint 2D-frequency estimation problem \cite{rai2025multi}. Nonetheless, the range-angle estimation in wideband XL MIMO-FMCW radar cannot be directly posed as a 2D frequency estimation problem \cite{rai2025low}. There are three major challenges in the range-angle estimation of XL MIMO-FMCW radar - (i) The range and angle terms are coupled with each other \cite{rai2025low}, and every point target is spread as a cluster in the range-angle domain. (ii) The maximum of each cluster corrseponding to the target is shifted from the actual coarse range-angle bin due to the SWE \cite{rai2025two}. (iii) The clusters corresponding to closely spaced targets can overlap with each other in the range-angle domain and reduce the target detectability.

In \cite{park2024spatial,wang2018spatial}, the compensation method for the RAC effect and SWE is proposed without considering the migration of range-angle bin from the maximum of the target cluster (these works directly consider the maximum of the cluster as the coarse range-angle bin estimate). Further, in \cite{rai2025two}, it is shown that at very high bandwidth, there is a significant migration of the range-angle bin from the target cluster maximum, and a two-stage rotation-based algorithm is proposed in which the correct coarse bin is find through rotation in the first-stage followed by fine-tuning around the correctly estimated range-angle bin in the second stage. Moreover, a compressive sensing (CS) based robust method is proposed in \cite{rai2026low} to estimate the signature of the XL MIMO-FMCW radar, which is free from the range-angle bin migration effect. However, each of the works mentioned above consider the separate clusters for each targets are present in the range-angle domain, which might not be always true for closely spaced targets and under severly high RAC effect. To the best of our knowledge there does not exist any work in the literature that discusses about the target detection and range-angle parameter estimation with overlapping target clusters for XL-MIMO FMCW radars. 

In this work, at first, we propose a generic decoupling-based 2D frequency estimation, which we further extend to detect the overlapping targets and estimate their range-angle parameters in XL-MIMO FMCW radar. Below, we give the relevant literature of existing decoupling-based methods applied to different systems for frequency and parameter estimation.

\subsection{Decoupling-based frequency estimation}

In \cite{wang2006novel}, authors propose a novel 2D frequency estimation method by several 1D frequnecy estimation processes for the data lengths being much larger in one dimeension than the other. However, in many practical applications it might not be the always the case. To handle this, there exist CS-based methods like atomic norm minimization (ANM) \cite{bhaskar2013atomic,chi2014compressive}, orthogonal mathcing pursuit (OMP) \cite{rai2025low1} etc. In \cite{zhang2019efficient}, a decoupled ANM-based approach is suggested over the vectorized vesrion of ANM in \cite{chi2014compressive} where, the run time complexity for a $22\times22$ grid is reduced from $734$ seconds to $1.5$ seconds (Fig. 1 in \cite{zhang2019efficient}). However, the run time complexity is still high for real time systems and will be even higher for the high measuremnt grids like XL-MIMO FMCW radar. Further, a manifold separation technique is applied for decoupled azimuth and elevation angle estimation in \cite{zhang2017decoupled}. However, due to the use of root-MUSIC method applied for 1D frequency estimation, the computational complexity of the proposed method is still of cubic order. A decoupling based idea is also implemented recently in \cite{hu2022decoupled} for multiple audio source localization in reverberant environments. None of the works mentioned above consider the spatial wideband effect. In \cite{weng2023wideband}, author consider the SWE for millimeter-Wave channel estimation and localization, where the estimation is carried out per-subcarrier wise first and then an additional pairing mechanism is used across the subcarriers. Moreover, in our previous work \cite{rai2023sparse}, we consider the low-index based decoupling method for sparse target detection in MIMO-OFDM systems, using the discrete Fourier transform (DFT) algorithm, which successfully deals with the overlapping path detection. However, the work in \cite{rai2023sparse} still faces the path pairing problem and limited DFT resolution. In this work, we propose a more generalized decoupling based 2D frequency estimation algorithm, which automatically pairs the frequency at different axis and is extendable to the 2D frequency estimation of wideband XL MIMO-FMCW radar systems. The main contribution of our work is as below.

\subsection{Contributions}

\begin{itemize}
    \item \textbf{Unified signal modeling for XL-MIMO FMCW radar:} We derive the intermediate frequency (IF) signal models for both conventional spatial narrowband MIMO-FMCW radar and spatial wideband XL-MIMO FMCW radar systems. For the spatial narrowband case, the received IF signal is shown to be a superposition of two-dimensional complex exponentials, enabling the formulation of joint range-angle estimation as a 2D frequency estimation problem. For the spatial wideband case, we derive the closed-form IF signal that explicitly captures the SWE and demonstrate the resulting distortions, namely range migration (delay squint), beam squint, and range-angle coupling effects.
    \item \textbf{Generic decoupling-based 2D frequency estimation framework:} We propose a generic decoupling-based approach for joint 2D frequency estimation, where the frequencies along one dimension are first estimated using a super-resolution 1D sparse recovery technique, followed by estimation along the second dimension conditioned on the first estimate. The proposed framework employs OMP-based super-resolution frequency estimation, enabling accurate recovery of closely spaced spectral components while automatically pairing the frequency components across dimensions, thereby avoiding the conventional pairing ambiguity encountered in many decoupled approaches.
    \item \textbf{Extension to spatial wideband XL-MIMO radar for overlapping target:} The proposed decoupling framework is further extended to spatial wideband XL-MIMO FMCW radar systems, where conventional 2D fast Fourier transform (FFT) and clustering-based approaches fail due to severe range-angle coupling and target cluster overlap. By exploiting the low-index property of the spatial wideband term, the proposed method first estimates the DoAs, compensates the SWE using the estimated angles, and subsequently estimates the corresponding ranges. This strategy enables reliable detection and separation of overlapping targets even under severe RAC conditions at large bandwidths and array apertures.
    \item \textbf{Comprehensive simulation-based validation:} Extensive simulations are conducted to evaluate the effectiveness of the proposed algorithm under both spatial narrowband and spatial wideband scenarios. The results demonstrate that the proposed method accurately detects multiple targets, correctly estimates their corresponding range-angle frequency pairs, and successfully resolves targets whose responses overlap in the range-angle domain due to spatial wideband effects. The proposed framework significantly improves target detectability compared to conventional 2D-FFT or clustering-based methods.
\end{itemize}

\section{System Model} In this section, we derive the system model for conventional MIMO-FMCW radar and wideband XL MIMO-FMCW radar separately and pose the range-angle estimation problem as a joint 2D-frequency estimation problem.

\begin{figure}
    \centering
    \includegraphics[width=0.95\linewidth]{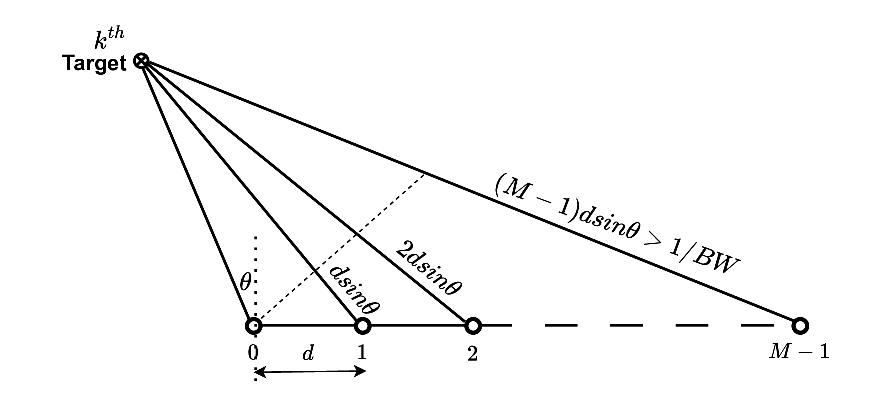}
    \caption{Uniform linear array corresponding to virtual elements of XL-MIMO radar}
    \label{fig:system_model}
\end{figure}

We consider a monostatic MIMO-FMCW radar with transmitters transmitting orthogonal linear frequency modulated (LFM) chirps of carrier frequency $f_c$, chirp rate $\gamma$, and chirp duration $T_{ch}$. The bandwidth of the chirp is denoted as $BW = \alpha f_c$, where $\alpha$ is the bandwidth selection parameter. The virtual linear array corresponding to the MIMO configuration is shown in Fig. \ref{fig:system_model}, which consists of $M$ virtual array elements, each placed with $d$ inter-element spacing. The radio scene consists of $K$ far-field sparse targets ($K\ll M$). We assume the static target scene and the $k^{th}$ located at $r_k$ distance at an elevation angle of $\theta_k$. The proposed framework can trivially be extended to dynamic targets as well.

\subsection{Conventional MIMO-FMCW Radar} 

The LFM chirp transmitted by a transmit antenna is modeled as
\par\noindent\small
\begin{align*}
s(t) = \exp{\left(j2\pi\left(f_{c}t+\frac{\gamma}{2}t^{2}\right)\right)}, \;\; 0\leq t\leq T_{ch}.
\end{align*}
\normalsize

We consider the time division multiplexing (TDM) radar so that all the transmitted chirps can be readily separated at the receivers. Accordingly, we first consider the received signal component at the $N_r$-th receiver for a single chirp transmitted from the $N_t$-th transmitter, given by
\par\noindent\small
\begin{align*}
r_{N_r,N_t}(t) = \sum_{k=1}^{K} a_{k} \, s(t-\tau^{k}_{N_r,N_t}), \;\; 0\leq t\leq T_{ch}.
\end{align*}
\normalsize
where $a_{k}$ is the complex amplitude proportional to the $k$-th target’s radar cross-section (RCS), and $\tau^{k}_{N_r,N_t}$ is the total time delay for the $k$-th target’s reflected signal. For the virtual array corresponding to the MIMO configuration as shown in Fig. \ref{fig:system_model}, the received signal at the $m$-th element of the virtual uniform linear array (ULA), can be written as
\par\noindent\small
\begin{align*}
r_{m}(t) = \sum_{k=1}^{K} a_{k} \, s(t-\tau^{k}_{m}), \;\; 0\leq t\leq T_{ch},
\end{align*}
\normalsize
where the total delay $\tau^{k}_{m}$ consists of a range delay, a Doppler-induced delay, and an angular delay, expressed as
\par\noindent\small
\begin{align}
\tau^{k}_{m} = \tau^{k}_{R} + \tau^{k}_{m,\theta}, \label{eqn:delay components}
\end{align}
\normalsize
with $\tau^{k}_{R}=2R_{k}/c$ and under the far-field assumption, $\tau^{k}_{q,\theta}\approx md\sin{(\theta_{k})}/c$.

At the $m$-th virtual receiver, the received signal $r_{m}(t)$ is mixed with the transmitted chirp $s(t)$ to obtain the IF signal as
\par\noindent\small
\begin{align*}
    y_{m}(t)=s(t)r^{*}_{m}(t)=\sum_{k=1}^{K}a_{k}^{*}&\exp{\left(j2\pi\gamma\tau^{k}_{m}t\right)} \exp{\left(-j\pi\gamma(\tau^{k}_{m})^{2}\right)}
    \nonumber\\
    &\;\;\times \exp{\left(j2\pi f_{c}\tau^{k}_{m}\right)},
\end{align*}
\normalsize
which is then sampled at sampling frequency $f_{s}$ to yield the (discrete) fast-time measurements
\par\noindent\small
\begin{align}
    y_{m}[n]=\sum_{k=1}^{K}a_{k}^{*}\underbrace{\exp{\left(j2\pi\gamma\tau^{k}_{m}\frac{n}{f_{s}}\right)}}_{\textrm{Term-I}}&
    \underbrace{\exp{\left(-j\pi\gamma(\tau^{k}_{m})^{2}\right)}}_{\textrm{Term-II}}\nonumber\\
    &\underbrace{\exp{\left(j2\pi f_{c}\tau^{k}_{m}\right)}}_{\textrm{Term-III}}.\label{eqn:discrete IF signal}
\end{align}
\normalsize

We, henceforth, deal with only discrete-time signals and use $n$ to denote the discrete-time index with $0\leq n\leq N-1$ where $N=\lceil f_{s}T_{ch}\rceil$. Further, we investigate the terms I, II, and III for a practical radar setup.

\textit{Term-I:} We define $\Omega_{R}^{k}\doteq\gamma \tau^R_k/f_s$ and normalized angular/spatial frequency $\Omega^{k}_{\theta}\doteq d\sin(\theta_{k})/\lambda$, $\gamma = BW/T_{ch}$. 
\par\noindent\small
\begin{align} 
    \exp{\left(j2\pi\gamma\tau_m^k \frac{n}{f_s}\right)} = &\exp{\left(j2\pi\frac{\gamma}{cf_s}(2R_k+md\sin(\theta_k))n\right)}.
    \label{eq:term_1}
\end{align}
\normalsize

For conventional MIMO-FMCW radar with a small array aperture, the spatial delay is much less than the radial delay (i.e., $\tau^k_{\theta} \ll \tau^k_R$). Hence, we can ignore the second term in \eqref{eq:term_1}. 

\par\noindent\small
\begin{align} 
    \exp{\left(j2\pi\gamma\tau_m^k \frac{n}{f_s}\right)} = \exp{\left(j2\pi\Omega_R^kn\right)}.
\end{align}
\normalsize

\textit{Term-II:} For practical radar system parameters \cite{lovescu2020fundamentals}, the quadratic term-II is usually ignored. Substituting \eqref{eqn:delay components} in Term-II we get
\par\noindent\small
\begin{align*}
    \exp{(-j\pi\gamma(\tau_m^k)^2)} &= \exp{\left(-j\pi\gamma \frac{4R_k^2}{c^2}\right)}\exp{\left(-j\pi\gamma\frac{m^2d^2\sin^2\theta_k}{c^2}\right)}\\
    &\exp{\left(-j2\pi\gamma\frac{2R_kmd\sin\theta_k}{c^2}\right)}.
\end{align*}
\normalsize
Here, $\exp{(-j\pi\gamma {4R_k^2}/{c^2})}$ does not vary with antenna index and hence can be absorbed in $\widetilde{a}_{k}$. Further, the maximum value of the remaining two terms in the exponents is close to zero and hence can be ignored.

\textit{Term-III:} Substituting \eqref{eqn:delay components} in Term-III, we obtain
\par\noindent\small
\begin{align*}
    \exp{\left(j2\pi f_{c}\tau^{k}_{m}\right)}=\exp{\left(j2\pi f_{c}\tau^{R}_{k}\right)}\exp{\left(j2\pi\frac{md\sin(\theta_{k})}{\lambda}\right)}.
\end{align*}
\normalsize
Here, $\exp{\left(j2\pi f_{c}\tau^{R}_{k}\right)}$ does not vary with antenna elements and hence, can be included in $\widetilde{a}_{k}$ in \eqref{eqn:2D_swb}. We can write the equivalent IF signal in \eqref{eqn:discrete IF signal} with spatial narrowband assumption as
\par\noindent\small
\begin{align}
    y_{m}[n]\approx\sum_{k=1}^{K}\widetilde{a}_{k}\exp{(j2\pi\Omega^{k}_{R}n)}\exp{(j2\pi\Omega^{k}_{\theta}m)}+w_{m}[n],\label{eqn:2D_snb}
\end{align}
\normalsize
where $\widetilde{a}_{k}=a_{k}^*exp(j\pi\gamma(\tau^{R}_{k})^{2})exp(-j2\pi f_{c}\tau^{R}_{k})$ and $w_m[n]$ is additive white Gaussian noise (AWGN) which follows $\mathcal{CN}(0,\sigma^2)$. It can be noticed that \eqref{eqn:2D_snb} is a mixture of $K$ 2D-complex tones, and hence the beat frequency ($\Omega^{k}_{R}$) and spatial frequency ($\Omega^{k}_{\theta}$) can be estimated with a joint 2D-frequency estimation algortihm.

\subsection{Wideband XL MIMO-FMCW Radar} For a wideband XL MIMO-FMCW radar, the spatial delay across the large array aperture cannot be ignored, and hence we have to consider the second term in \eqref{eq:term_1}. With this the term-I can be equivalently written as

\par\noindent\small
\begin{align} 
    \exp{\left(j2\pi\gamma\tau_q^k \frac{n}{f_s}\right)} = &\exp{\left(j2\pi\frac{\gamma}{cf_s}(2R_k+qd\sin(\theta_k))n\right)}\nonumber\\
    &= \exp{\left(j2\pi\Omega_R^kn\right)}\exp{\left(j2\pi\frac{\Omega^k_{\theta}}{f_c}\gamma\frac{mn}{f_s}\right)}\nonumber\\ &=\exp{\left(j2\pi\Omega_R^kn\right)}\exp{\left(j2\pi\Omega^k_{\theta}\frac{\alpha}{N}mn\right)}.
    \label{eq:term_1_swb}
\end{align}
\normalsize

Using \eqref{eq:term_1_swb}, the IF signal for XL MIMO-FMCW radar with spatial wideband assumption can be written as 

\par\noindent\small
\begin{align}
    y_{m}[n]\approx\sum_{k=1}^{K}\widetilde{a}_{k}&\exp{(j2\pi\Omega^{k}_{R}n)}\exp{(j2\pi\Omega^{k}_{\theta}m)}\nonumber\\
    &\;\;\times \underbrace{\exp{(j2\pi\frac{\alpha}{N}\Omega^{k}_{\theta}mn)}}_{SW\;Term}+w_{m}[n],\label{eqn:2D_swb}.
\end{align}
\normalsize

It should be noted here that, unlike \eqref{eqn:2D_snb}, in \eqref{eqn:2D_swb} there is an extra spatial wideband (SW) term which is coupled in space-time index, which raises the RAC effect in wideband XL MIMO-FMCW radar. The coupling effect in the SW-term here is dependent upon three parameters - (i) DoA, (ii) BW selection parameter, and (iii) number of array elements. For smaller BWs, $\alpha \rightarrow 0$, the SW term is approximated by 1, and the equivalent MIMO-FMCW IF signal becomes a mixture of 2D complex exponential tones, whereas for large BWs, the IF signal does not resemble a complex exponential tone mixture and hence new algorithms are required to handle this situation. In this work, we propose a generic decoupling-based 2D-frequency estimation algorithm in which the frequencies along one dimensions are estimated first using 1D-frequency estimation, and then for every estimated frequency in the first dimension, we estimate the frequencies along the other dimension again using 1D-frequency estimation. Our proposed technique automatically pairs the estimated frequencies along the two separate dimensions. Furthermore, the two-tuple frequencies can be well separated in a 2D plane, whereas the corresponding 1D frequencies can be closely spaced, which might not be possible to be estimated by directly applying methods like discrete Frequency transform (DFT). In this work, we propose the use of CS-based super-resolution 1D frequency estimation method using the OMP algorithm.

\section{Spatial Wideband Effect in MIMO-FMCW Radar}

\begin{figure*}
\centering
\begin{subfigure}{0.30\textwidth}
    \includegraphics[width=\textwidth]{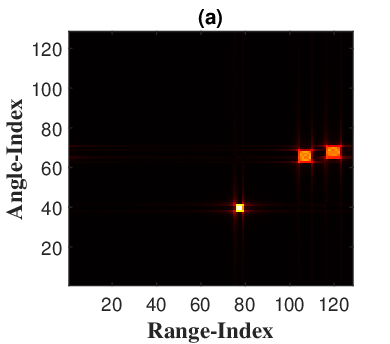}
\end{subfigure}
\hfill
\begin{subfigure}{0.30\textwidth}
    \includegraphics[width=\textwidth]{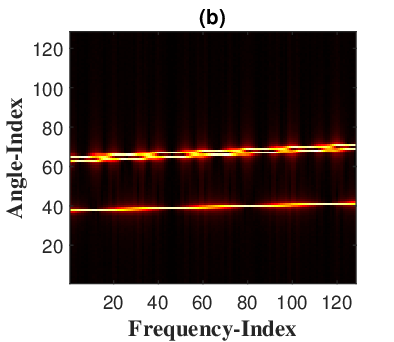}
\end{subfigure}
\hfill
\begin{subfigure}{0.30\textwidth}
    \includegraphics[width=\textwidth]{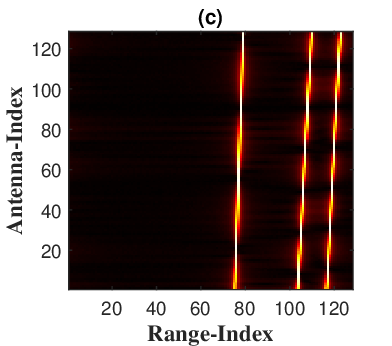}
\end{subfigure}
\caption{Illustration of signal distortions in XL-MIMO FMCW radar systems. (a) range–angle map showing dual wideband spread due to RAC. (b) frequency-dependent beam squint effect (c) range migration effect across the large antenna array.}
\label{fig:swe_radar}
\end{figure*}

We can write the system model in \eqref{eqn:2D_swb}, in the matrix form as following 
\par\noindent\small
\begin{align}
       \mathbf{Y} = \sum\limits_{k=1}^{K} \tilde{a}_k \mathbf{b}(\Omega_{\theta}^k) \mathbf{a}^T(\Omega_R^k) \circ \mathbf{S}(\alpha,\Omega_{\theta}^k),
    \label{eqn:spatial_wideband_frequency_wideband}
\end{align}
\normalsize

where, $\mathbf{b}(\Omega_{\theta}^k) \doteq [1,\cdot \cdot\cdot, \exp(j2\pi \Omega_{\theta}^k(M-1))]^T$, $\mathbf{a}(\Omega_{R}^k) \doteq [1,\cdot\cdot\cdot, \exp(j2\pi \Omega_{R}^k(N-1))]^T$, and $[\mathbf{S}]_{m,n} \doteq \exp(j2\pi \alpha \Omega_{\theta}^k/N\cdot mn)$ are the angle steering vector, range steering vector, and wideband phase shift matrix, respectively.

\subsection{Beam-Squint Effect} 
We can transform the IF signal from the space-time domain to the angle-time domain via taking DFT w.r.t. the space domain ($m$). In the spatial wideband MIMO systems, this cannot be directly utilized for angle estimation as the angular response is not constant for the entire BW range. Mathematically, for a $k^{th}$ path, we take the normalized DFT of \eqref{eqn:2D_swb} w.r.t. antenna-index $m$ as

\par\noindent\small
\begin{align}
    \mathbf{Z}_k[p,n] &= \frac{1}{\sqrt{M}}\tilde{a}_k\exp(j2\pi\Omega_R^k) \times \nonumber\\ & \sum_{m=0}^{M-1} \exp\left(j2\pi \Omega_{\theta}^k\left(1+\frac{\alpha n}{N}\right)\right)\exp\left(j\frac{2\pi}{M}mp\right)\nonumber\\
    & = \tilde{a}_k\exp(j2\pi\Omega_R^k) \mathcal{D}\left
    (\Omega_{\theta}^{k}\left(1+\frac{\alpha n}{N}\right)-\frac{p}{M}\right),
\end{align}
\normalsize

where, $(1/\sqrt{L})\sum_{l=0}^{L-1}\exp(j2\pi lx) = \sqrt{Q}\exp(j2\pi x(L-1)) \sin\pi Lx/\sin\pi x \doteq \mathcal{D}_L(x)$ and in the asymptotic case $\lim_{L\rightarrow\infty}|\mathcal{D}_L(x)| = \sqrt{L}\delta(x)$.

\subsection{Delay-Squint Effect} In MIMO-FMCW radar, the delay term gives rise to the normalized beat frequency and hence corresponds to the range of a target. The delay squint effect is also popularly known as the range migration effect. To investigate this effect, we take the normalized DFT of \eqref{eqn:2D_swb} w.r.t. fast time-index ($n$) as

\par\noindent\small
\begin{align}
    \mathbf{Z}_k[m,q] &= \frac{1}{\sqrt{N}}\tilde{a}_k\exp(j2\pi\Omega_{\theta}^k)  \nonumber\\ & \times \sum_{n=0}^{N-1} \exp\left(j2\pi \left(\Omega_{R}^k+\frac{\alpha n\Omega_{\theta}^k}{N}\right)\right)\exp\left(j\frac{2\pi}{N}nq\right)\nonumber\\
    & = \tilde{a}_k\exp(j2\pi\Omega_{\theta}^k) \mathcal{D}\left
    (\left(\Omega_{R}^{k}+\frac{\alpha n\Omega_{\theta}^k}{N}\right)-\frac{q}{N}\right).
\end{align}
\normalsize

\subsection{Dual Wideband Effect} In the angle-range domain, the channel response corresponding to a particular target is spread equally in both domains simultaneously and hence is a.k.a. range-angle coupling effect. This can be observed by taking the 2D-DFT of \eqref{eqn:2D_swb} as

\par\noindent\small
\begin{align}
    \mathbf{G}_k(p,q) &= \frac{1}{\sqrt{MN}} \sum_{m=0}^{M-1}\sum_{n=0}^{N-1} \tilde{a}_k \exp(j2\pi\Omega_{\theta}^{k}m)\exp(j2\pi\Omega_{R}^{k})\nonumber\\
    & \underbrace{\exp\left(j2\pi\frac{\alpha}{N}mn\right)}_{SW\;Term} \times \exp\left(j\frac{2\pi}{M}mp\right) \exp\left(j\frac{2\pi}{N}nq\right).
    \label{eqn:inter_dwb}
\end{align}
\normalsize

It should be noticed that due to the presence of the range-angle coupling term, \eqref{eqn:inter_dwb} cannot be directly solved in closed form. Hence, we first club the SW term with the space index and take the summation to write the following 

\par\noindent\small
\begin{align}
    \mathbf{G}_k(p,q) = \frac{1}{\sqrt{N}} \sum_{n=0}^{N-1} & \tilde{a}_k \exp(j2\pi\Omega_{R}^{k}n)\exp\left(j\frac{2\pi}{N}nq\right)\nonumber \nonumber\\
    & \mathcal{D}\left
    (\Omega_{\theta}^{k}\left(1+\frac{\alpha n}{N}\right)-\frac{p}{M}\right).
    \label{eqn:inter_swb1}
\end{align}
\normalsize

Next, we combine the SW term with the time-index and can write the following

\par\noindent\small
\begin{align}
    \mathbf{G}_k(p,q) = \frac{1}{\sqrt{M}} \sum_{m=0}^{M-1} & \tilde{a}_k \exp(j2\pi\Omega_{\theta}^{k}m)\exp\left(j\frac{2\pi}{M}mp\right)\nonumber\\
    & \mathcal{D}\left
    (\left(\Omega_{R}^{k}+\frac{\alpha n\Omega_{\theta}^k}{N}\right)-\frac{q}{N}\right).
    \label{eqn:inter_swb2}
\end{align}
\normalsize

Now, combining \eqref{eqn:inter_swb1} and \eqref{eqn:inter_swb2}, we can observe that in the angle-range domain, the amount of spread is equal in both dimensions.

We demonstrate the above-mentioned key challenges associated with XL-MIMO FMCW radar operation in Fig. \ref{fig:swe_radar}. As shown in Fig. \ref{fig:swe_radar}(a), the range–angle map exhibits significant spreading jointly in the range-angle domain, which stems from SWE introduced by the very large aperture. Fig. \ref{fig:swe_radar}(b) shows the range–frequency response, where pronounced frequency-dependent beam squint and spreading become evident due to the use of ultra-wide bandwidth that results in degraded angular resolution and target localization accuracy. Moreover, Fig. \ref{fig:swe_radar}(c) depicts the range–antenna domain, where the range bins are migrated sufficiently, which arises as the antenna aperture grows large relative to the operating wavelength. These artifacts underscore the drawbacks of scaling FMCW radar systems to extreme array sizes and bandwidths, motivating the need for novel low-complexity algorithms to compensate SWE and related distortions for precise parameter estimation.

\section{Preliminary: OMP-based frequency estimation}
\label{sec:omp}
In this section, we give a primer on frequency estimation using an OMP-based super-resolution method over the conventional DFT-based method. For a complex exponential mixture of $K$ tones, the signal model can be written as 
\par\noindent\small
\begin{align}
    y[n] = \sum_{k=1}^{K} a_k \exp(j2\pi f_k n) + w[n], \forall n = 0\hdots N-1,
    \label{eq:1d}
\end{align}
\normalsize

where, $w[n]$ is complex Gaussian noise which follows $\mathcal{CN}(0,\sigma^2)$. Now, we define the $K\times 1$ vector $\Tilde{x} = [x_{1},\hdots,x_{K}]^{T}$ such that \eqref{eq:1d} can be written as 
\par\noindent\small
\begin{align}
    \mathbf{y} = \tilde{\mathbf{A}}(\mathbf{f})\tilde{x} +\mathbf{w},
    \label{eq:omp_1d}
\end{align}

where the $N\times K$ matrix $\widetilde{\mathbf{A}}(\mathbf{f})=[\mathbf{a}(f^{1}),\hdots,\mathbf{a}(f^{k})]$ with the $k$-th column $\mathbf{a}(f^{k})\\\doteq [1,\exp{(j2\pi f^{k})},\hdots,\exp{(j2\pi f^{k} (N-1))}]^{T}$. Here, $\mathbf{w}$ represents the $N\times 1$ stacked noise vector. Each column $\mathbf{a}(f)$ of matrix $\widetilde{\mathbf{A}}$ is parameterized by the frequency $f$. Furthermore, $\mathbf{y}$ in \eqref{eq:omp_1d} represents full measurements in the time domain sampled at frequency $f_{s}$.

We choose a grid of $G$ points $\{\omega_{g}:1\leq g\leq G\}$ of the possible frequencies $f$ with $G \gg K$ and negligible discretization errors. Substituting these grid points in $\mathbf{a}(\cdot)$, we construct an over-complete $N\times G$ measurement matrix $\mathbf{A}=[\mathbf{a}(f),\hdots,\mathbf{a}(\omega_{G})]$. Then, \eqref{eq:omp_1d} becomes
\par\noindent\small
\begin{align}   \mathbf{y}=\mathbf{A}\mathbf{x}+\mathbf{w},\label{eqn:omp_1d_final}
\end{align}
\normalsize
where the $G\times 1$ vector $\mathbf{x}$ contains the target ranges and unknown coefficients $\{x_{k}\}$. In particular, a non-zero element of $\mathbf{x}$ represents a target present at the range corresponding to the grid point. Since $K\ll G$, $\mathbf{x}$ is a sparse vector and frequency estimation reduces to determining $\textrm{supp}(\mathbf{x})$ given $\mathbf{y}$ and $\mathbf{A}$. Hence, we can utilize 1D sparse recovery algorithms (like matching pursuit or basis pursuit) to recover the $supp(\mathbf{x})$. Owing to its low computational complexity, in this work, we propose to use the OMP algorithm to recover the target frequencies. 

\section{Decoupling-based 2D Frequency Estimation}
\label{sec:dec}
In this section, we propose a generic decoupling-based estimation of joint 2D-frequency estimation while addressing the conventional narrowband MIMO-FMCW radar's range-angle estimation. In the range-angle domain, there may be a few targets that have the same range with different DoAs; let there be a total $R \leq K$ number of distinct ranges of targets. In such a condition, these few range paths may have more than one angular path. Hence, for a particular range $r$ the number of angular paths are $S^{r}$ such that $K =  \sum_{r=1}^{R} S^r$. If we represent the beat frequency of the $r$-th range path with $\Omega^{r}_{R}$, and the angular frequency of the $s^{r}$-th
angular path for the $r$-th range path with $\Omega^{s^r}_{\theta}$
then the signature of a $k$-th target is written as $(\Omega_{R}^{k}, \Omega_{\theta}^{k}, \tilde{a}^k)$ $\equiv$ 
$(\Omega_{R}^{r}, \Omega_{\theta}^{s^r}, \tilde{a}^{s^r})$. Hence, we can rewrite \eqref{eqn:2D_snb} as 

\par\noindent\small
\begin{align}
    [\mathbf{Y}]_{m,n} = \sum\limits_{r=1}^{R}\exp(j2\pi\Omega_R^rn)\sum_{s^r=1}^{S^r}\tilde{a}^{s^r}\exp(j2\pi\Omega_{\theta}^{s^r}m),
\end{align}
\normalsize
where, $\mathbf{Y}$ is stacked with the measurement of space-indices across the row and time-indices across the column. Now, for a particular antenna-index, we can write the time domain-only signal from \eqref{eqn:2D_snb} as 

\par\noindent\small
\begin{align}
    \mathbf{y}[n]\big|_{\!m} = \sum_{r=1}^{R} \tilde{b}_{r}(m)\exp(j2\pi\Omega_{R}^{r}n),
    \label{eqn:time_1D}
\end{align}
\normalsize
where $\tilde{b}_{r}(m) = \sum_{s=1}^{S^r}\tilde{a}^{s^r}\exp(j2\pi\Omega_{\theta}^{s^r}m)$, is the effective complex path coefficient for a given antenna-index. We can take the 1D normalized DFT of \eqref{eqn:2D_snb} and write

\par\noindent\small
\begin{align}
    \mathbf{y}[q]\big|_{\!m}= \frac{1}{\sqrt{R}}\sum\limits_{r=1}^{R}\tilde{b}_{r}(m)\mathcal{D}\left(\Omega_{R}^{r}-\frac{q}{N}\right).
    \label{eqn:range_1D}
 \end{align}
\small

At a given antenna index, we can find all the targets separable in the range bins. Let us calculate the response of \eqref{eqn:range_1D} at the $0^{th}$ antenna-index, where $\tilde{b}_{r}(0) = \sum_{s=1}^{S^r}\tilde{a}^{s^r}$. In the asymptotic case $\lim n\rightarrow \infty$, we can write
\par\noindent\small
\begin{align}
     \mathbf{y}[q]\big|_{\!0}= \frac{1}{\sqrt{R}}\sum\limits_{r=1}^{R}\tilde{b}_{r}(0)\delta\left(\Omega_{R}^{r}-\frac{q}{N}\right).
    \label{eqn:range_1D_asymp}
\end{align}
\normalsize

However, with a limited number of measurements along fast time, the range resolution is limited to $1/N$ and the accuracy of the estimated beat-frequency and corresponding range is also limited. Instead, knowing that the observed radio scene is sparse, we can cast the problem of targets' range estimation in \eqref{eqn:time_1D} as a sparse recovery problem, suggested in Section~\ref{sec:omp}, and achieve the super-resolution with high accuracy.

Now, we take the DFT across each column of $\mathbf{Y}$ and stack them in an intermediate range-antenna matrix $\mathbf{\tilde{Y}}$. Again, exploiting the sparsity of the observed radio scene, we need not search across all the range bins, but rather search only those range bins where the targets are present, which further reduces the computational complexity. Hence, for a particular range bin, we can write the space domain vector from $\mathbf{\tilde{Y}}$ as
\par\noindent\small
\begin{align}
    \mathbf{\tilde{y}}[m]\big|_{\!q} = \sum\limits_{s^r=1}^{S^r} \underbrace{\tilde{a}^{s^r} \mathcal{D}\left(\Omega_{R}^{r}-\frac{q}{N}\right)}_{\tilde{c}_{\theta}(q)} \exp(j2\pi\Omega_{\theta}^{s^r}m). 
    \label{eqn:angle_1D}
\end{align}
\normalsize

Similar to above, we can either take the 1D normalized DFT or use the sparse recovery algorithm from Section~\ref{sec:omp} for estimating the target angles at a particular range bin from \eqref{eqn:angle_1D}. We can write the DFT of \eqref{eqn:angle_1D} as 
\par\noindent\small
\begin{align}
    \mathbf{\tilde{y}}[p]\big|_{\!q} &= \frac{1}{\sqrt{M}}\sum\limits_{s^r=1}^{S^r} \tilde{c}_{\theta}(q) \mathcal{D}\left(\Omega_{\theta}^{s^r}-\frac{p}{M}\right)\nonumber\\
    &=\lim_{M\rightarrow\infty}\sum\limits_{s^r=1}^{S^r} \tilde{c}_{\theta}(q) \delta\left(\Omega_{\theta}^{s^r}-\frac{p}{M}\right).
\end{align}
\normalsize

\begin{figure*}
\centering
\begin{subfigure}{0.23\textwidth}
    \includegraphics[width=\textwidth]{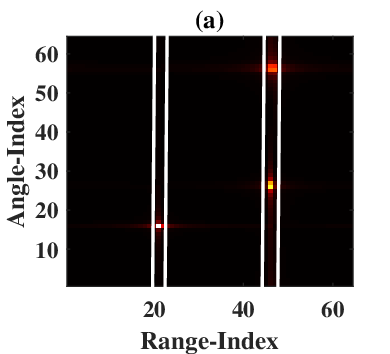}
\end{subfigure}
\hfill
\begin{subfigure}{0.23\textwidth}
    \includegraphics[width=\textwidth]{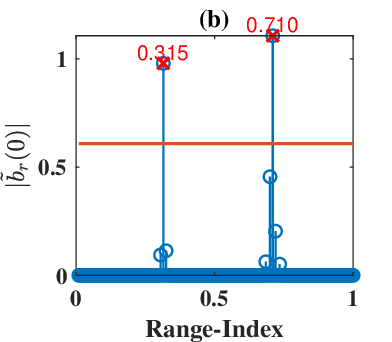}
\end{subfigure}
\hfill
\begin{subfigure}{0.23\textwidth}
    \includegraphics[width=\textwidth]{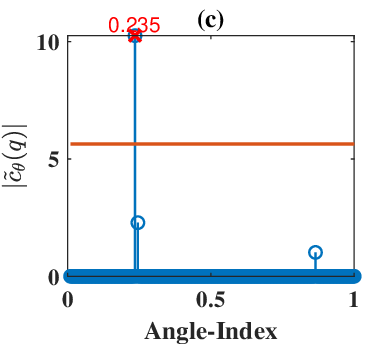}
\end{subfigure}
\hfill
\begin{subfigure}{0.23\textwidth}
    \includegraphics[width=\textwidth]{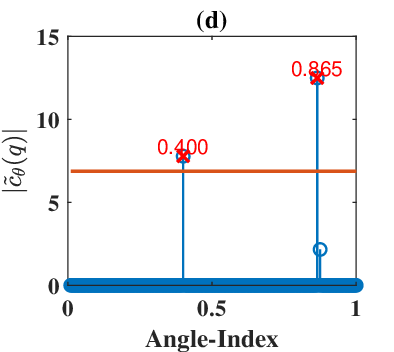}
\end{subfigure}
\caption{Illustration of proposed decoupling-based signature estimation (a) joint range–angle map with three targets across two range bins (b) range profile revealing two resolvable bins (c) angular profile at the first bin with a single target (d) angular profile at the second bin with two targets.}
\label{fig:decoupling_illustration}
\end{figure*}

\subsection*{Numerical Example} 

In Fig.~\ref{fig:decoupling_illustration}, we illustrate the proposed decoupling-based signature estimation strategy. 
In Fig.~\ref{fig:decoupling_illustration}(a), the joint range--angle map shows three targets located at normalized frequency pairs 
$(15.15/M,\,20.25/N)$, $(25.45/M,\,45.15/N)$, and $(55.45/M,\,45.50/N)$. 
Notably, the last two targets lie in the same range bin but at different angles, making their separation 
challenging in conventional approaches. 
Fig.~\ref{fig:decoupling_illustration}(b) shows the normalized range correlation function, which successfully resolves two distinct range bins. Figs.~\ref{fig:decoupling_illustration}(c) and (d) present the angular correlation profiles for each bin- a single angular target is detected in the first range bin, while two angular targets are clearly resolved in the second. These results confirm that the proposed method, leveraging an overcomplete OMP dictionary, accurately detects and separates multiple targets under the given setting with high precision. Moreover, it should be noted that with the proposed technique, the ranges and angles of the detected targets are automatically coupled, without any additional coupling mechanism required.

\section{Low-Index based Detection for XL-MIMO FMCW Radar}

\begin{figure}
    \centering
    
    \begin{subfigure}[t]{0.48\columnwidth}
        \centering
        \includegraphics[width=\linewidth]{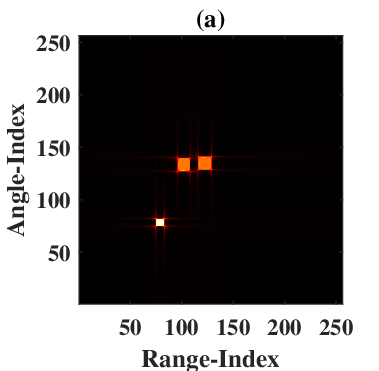}
    \end{subfigure}
    \hfill
    \begin{subfigure}[t]{0.48\columnwidth}
        \centering
        \includegraphics[width=\linewidth]{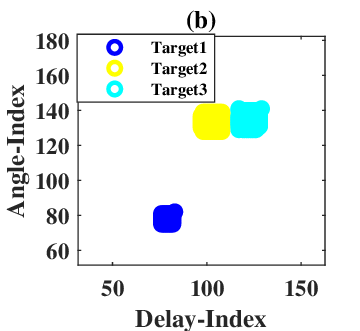}
    \end{subfigure}
    
    \begin{subfigure}[t]{0.48\columnwidth}
        \centering
        \includegraphics[width=\linewidth]{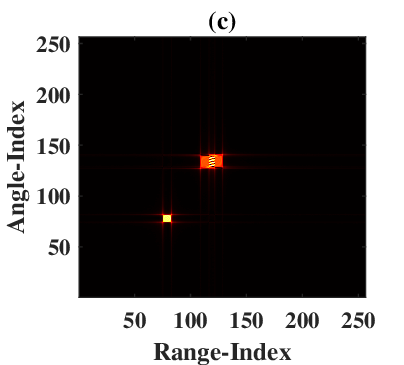}
    \end{subfigure}
    \hfill
    \begin{subfigure}[t]{0.48\columnwidth}
        \centering
        \includegraphics[width=\linewidth]{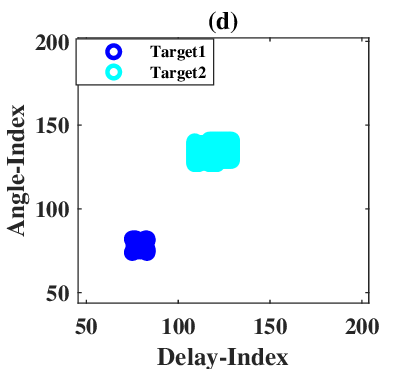}
    \end{subfigure}
    
    \caption{Range–angle maps and clustering results for three targets under different configurations, demonstrating the impact of SWE on target separability.}
    \label{fig:range_angle_clustering}
\end{figure}

In this section, we first illustrate the problem of target overlapping due to SWE in XL-MIMO FMCW radar.
At first, we simulate the range–angle map for a $256 \times 256$ measurement grid with three targets located at $(1.45\:\text{m},35^\circ)$, $(1.85\:\text{m},82^\circ)$, and $(2.25\:\text{m},85^\circ)$, as shown in Fig.~\ref{fig:range_angle_clustering}(a). In this case, although the targets are closely spaced, the local gravitation-based clustering (LGC) algorithm \cite{wang2017clustering} is able to successfully identify all three clusters corresponding to the true target locations, as illustrated in Fig.~\ref{fig:range_angle_clustering}(b).  
Next, we shift the second target to $2.10\:\text{m}$ while keeping the others fixed. As shown in Fig.~\ref{fig:range_angle_clustering}(c), even though the two targets are theoretically separable beyond the resolution limit, the SWE causes their spreads to overlap in the range–angle map. Consequently, the clustering outcome in Fig.~\ref{fig:range_angle_clustering}(d) reveals that only two clusters are detected, with one target completely missed. This demonstrates that under overlapping conditions induced by SWE, conventional clustering-based (or 2D-peak finding) approaches inevitably fail to resolve all targets.  

To this end, we propose a low-index decoupling-based approach for signature estimation of XL-MIMO FMCW radar, and we show that our proposed algorithm is able to resolve the overlapping targets. Moreover, unlike the narrowband case, where we could have approached the algorithm from either direction, in the wideband case, we have to start with the space dimension first because we need to compensate for the SWE from the estimated angle estimates. Hence, we can write the \eqref{eqn:2D_swb} in the decoupled way as

\par\noindent\small
\begin{align}
    [\mathbf{Y}]_{m,n} &= \sum\limits_{s=1}^{S}\exp(j2\pi\Omega_{\theta}^sm)\exp\left(j2\pi\frac{\alpha}{N}\Omega_{\theta}^{s}mn\right)\nonumber\\
&~~~~\sum_{r=1}^{R^s}\tilde{a}^{r^s}\exp(j2\pi\Omega_{R}^{r^s}n)
\end{align}
\normalsize

Further, for a particular time index, we can write the space-domain only signal from \eqref{eqn:2D_swb} as
\par\noindent\small
\begin{align}
    \mathbf{y}[m]\big|_{\!n} = \sum_{s=1}^{S} \tilde{d}_{s}(n)\exp(j2\pi\Omega_{\theta}^{s}m)\exp\left(j2\pi\frac{\alpha}{N}\Omega_{R}^{r}mn\right),
    \label{eqn:time_1D_swb}
\end{align}
\normalsize

where $\tilde{d}_{s}(n) = \sum_{r=1}^{R^s}\tilde{a}^{r^s}\exp(j2\pi\Omega_{R}^{r^s}n)$ is the effective complex path coefficient for a given time-index.

Now, it should be noted that for the low-time index, the effect of SW term in \eqref{eqn:time_1D_swb} can be neglected, i.e. at $n=0$, $\exp(j2\pi(\alpha/N)\Omega_{\theta}mn) = 1$. 
We can write the equivalent 1D complex tone mixture in the space domain as
\par\noindent\small
\begin{align}
    \mathbf{y}[m]\big|_{\!n} = \sum_{s=1}^{S} \tilde{d}_{s}(n)\exp(j2\pi\Omega_{\theta}^{s}m).
    \label{eqn:time_low_1D_swb}
\end{align}
\normalsize

\begin{figure*}
\centering
\begin{subfigure}{0.23\textwidth}
    \includegraphics[width=\textwidth]{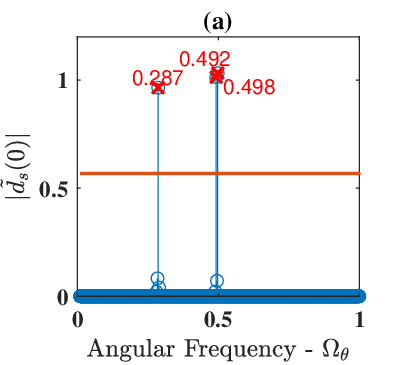}
\end{subfigure}
\hfill
\begin{subfigure}{0.23\textwidth}
    \includegraphics[width=\textwidth]{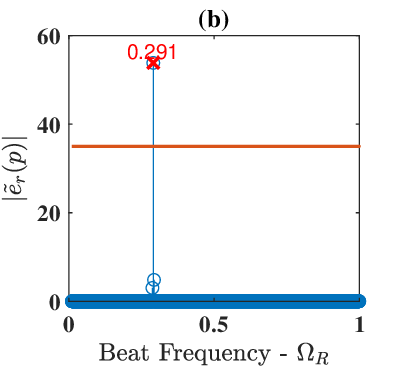}
\end{subfigure}
\hfill
\begin{subfigure}{0.23\textwidth}
    \includegraphics[width=\textwidth]{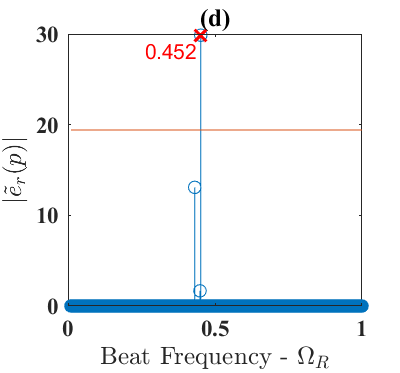}
\end{subfigure}
\hfill
\begin{subfigure}{0.23\textwidth}
    \includegraphics[width=\textwidth]{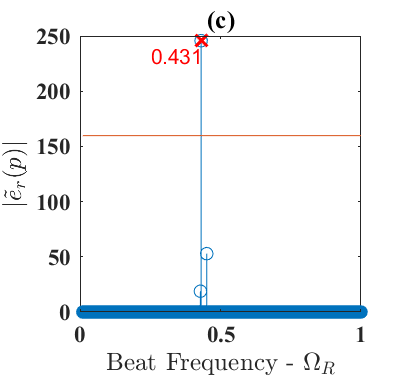}
\end{subfigure}
\caption{Detection of targets in the range--angle domain using the proposed decoupling-based low-index method
(a) angular frequency domain response showing successful detection of all target DoAs. 
(b)--(d) corresponding beat frequency responses for each detected DoA, clearly separating the overlapping targets in the range domain.}
\label{fig:decoupling_swb_illustration}
\end{figure*}

Now we can use either the 1D-DFT or the OMP-based super-resolution method to find the required target angles. For the next step, we first remove the SWE using the estimated angle via 1D-OMP for a particular target.

\par\noindent\small
\begin{align}
    [\mathbf{Y}]_{m,n} &= \sum\limits_{s=1}^{S}\exp(j2\pi\Omega_{\theta}^sm)\underbrace{\exp\left(j2\pi\frac{\alpha}{N}(\Omega_{\theta}^{s}-\hat{\Omega}_{\theta}^s)mn\right)}_{Residue ~Term}\nonumber\\
&~~~~\sum_{r=1}^{R^s}\tilde{a}^{r^s}\exp(j2\pi\Omega_{R}^{r^s}n)
\end{align}
\normalsize

For a closely estimated $\hat{\Omega}_{\theta}^{s}$, the $Residue~Term\rightarrow1$ and the SWE can be compensated for that particular target. Therefore, we can write the equivalent IF signal matrix as
\par\noindent\small
\begin{align}
    [\mathbf{\check{Y}}]_{m,n} &= \sum\limits_{s=1}^{S}\exp(j2\pi\Omega_{\theta}^sm) \sum_{r=1}^{R^s}\tilde{a}^{r^s}\exp(j2\pi\Omega_{R}^{r^s}n)
    \label{eqn:compensated_swb}
\end{align}
\normalsize

Furthermore, we construct the intermediate angle-time matrix by taking the 1D-DFT along the columns of $\mathbf{\check{Y}}$. For a particular angle bin, we can write the time domain vector from $\mathbf{\check{Y}}$ as
\par\noindent\small
\begin{align}
    \mathbf{\check{y}}[n]|_p = \sum\limits_{r=1}^{R^s}\underbrace{\tilde{a}^{r^s}\mathcal{D}\left(\Omega_{\theta}^{s}-\frac{p}{M}\right)}_{\tilde{e}_{r}(p)}\exp(j2\pi\Omega_{R}^{r^s}n).
    \label{eqn:1D_time_swb}
\end{align}
\normalsize

We can write the DFT of \eqref{eqn:1D_time_swb} as
\par\noindent\small
\begin{align}
    \mathbf{\check{y}}[q]|_p =\frac{1}{\sqrt{N}}\sum\limits_{r^s = 1}
^{R^s}\tilde{e}_{R}(p)\mathcal{D}\left(\Omega_{R}^{r^s}-\frac{q}{N}\right)\nonumber\\
\lim_{N\rightarrow\infty}\sum\limits_{r^s = 1}
^{R^s}\tilde{e}_{R}(p)\delta\left(\Omega_{R}^{r^s}-\frac{q}{N}\right).
\end{align}
\normalsize

As shown in Fig.~\ref{fig:range_angle_clustering}(c), targets 2 and 3 overlap in the range-angle domain, making it difficult to distinguish them directly.  To overcome this, our proposed method first estimates the directions of arrival (DoAs) at the $0$-th time index. From Fig.~\ref{fig:decoupling_swb_illustration}(a), it is evident that the OMP-based super-resolution technique can accurately resolve all target DoAs. 
For each detected angle, the corresponding range is then estimated, as depicted in Fig.~\ref{fig:decoupling_swb_illustration}(b)--(d). This joint angle--range coupling enables precise separation and localization of the targets. For instance, given the input frequencies of $(0.2868, 0.2908)$, $(0.4924, 0.4211)$, and $(0.4981, 0.4512)$, the detected frequency pairs corresponding to the three targets are $(0.2870, 0.2905)$, $(0.4925, 0.4310)$, and $(0.4980, 0.4515)$, respectively, demonstrating the high accuracy of the proposed approach. 
\vspace{0.5cm}
\paragraph*{Note:} A detailed numerical analysis of the proposed method w.r.t. different varying system parameters is being carried out and soon the current version will be updated.

\section{Conclusion}
In this work, we investigated the problem of target detection and joint range-angle estimation in spatial narrowband and spatial wideband XL-MIMO FMCW radar systems. We first derived the IF signal models for both cases and showed that while the spatial narrowband system naturally leads to a joint 2D frequency estimation problem, the spatial wideband case introduces spatial wideband effects such as beam squint, range migration, and range-angle coupling, which significantly distort the range-angle map. To address these challenges, we proposed a generic decoupling-based 2D frequency estimation framework that sequentially estimates the frequencies along two dimensions using super-resolution OMP-based sparse recovery. The proposed approach was further extended to spatial wideband XL-MIMO radar by exploiting the low-index property of the spatial wideband term, enabling compensation of the coupling effect and reliable estimation of target parameters. Simulation results demonstrated that the proposed method accurately detects targets and resolves overlapping target responses caused by range-angle coupling, outperforming conventional approaches. These results highlight the effectiveness of the proposed decoupling-based framework for high-resolution target detection in wideband XL-MIMO FMCW radar systems.

\ifCLASSOPTIONcaptionsoff
\newpage
\fi
\bibliographystyle{IEEEtran}
\bibliography{references}

\begin{thebibliography}{10}
\providecommand{\url}[1]{#1}
\csname url@samestyle\endcsname
\providecommand{\newblock}{\relax}
\providecommand{\bibinfo}[2]{#2}
\providecommand{\BIBentrySTDinterwordspacing}{\spaceskip=0pt\relax}
\providecommand{\BIBentryALTinterwordstretchfactor}{4}
\providecommand{\BIBentryALTinterwordspacing}{\spaceskip=\fontdimen2\font plus
\BIBentryALTinterwordstretchfactor\fontdimen3\font minus
  \fontdimen4\font\relax}
\providecommand{\BIBforeignlanguage}[2]{{%
\expandafter\ifx\csname l@#1\endcsname\relax
\typeout{** WARNING: IEEEtran.bst: No hyphenation pattern has been}%
\typeout{** loaded for the language `#1'. Using the pattern for}%
\typeout{** the default language instead.}%
\else
\language=\csname l@#1\endcsname
\fi
#2}}
\providecommand{\BIBdecl}{\relax}
\BIBdecl

\bibitem{hasch2012millimeter}
J.~Hasch, E.~Topak, R.~Schnabel, T.~Zwick, R.~Weigel, and C.~Waldschmidt,
  ``Millimeter-wave technology for automotive radar sensors in the 77 ghz
  frequency band,'' \emph{IEEE transactions on microwave theory and
  techniques}, vol.~60, no.~3, pp. 845--860, 2012.

\bibitem{patole2017automotive}
S.~M. Patole, M.~Torlak, D.~Wang, and M.~Ali, ``Automotive radars: A review of
  signal processing techniques,'' \emph{IEEE Signal Processing Magazine},
  vol.~34, no.~2, pp. 22--35, 2017.

\bibitem{waldschmidt2021automotive}
C.~Waldschmidt, J.~Hasch, and W.~Menzel, ``Automotive radar—from first
  efforts to future systems,'' \emph{IEEE Journal of Microwaves}, vol.~1,
  no.~1, pp. 135--148, 2021.

\bibitem{zhang2021mrpt}
Z.~Zhang, X.~Wang, D.~Huang, X.~Fang, M.~Zhou, and Y.~Zhang, ``Mrpt:
  Millimeter-wave radar-based pedestrian trajectory tracking for autonomous
  urban driving,'' \emph{IEEE Transactions on Instrumentation and Measurement},
  vol.~71, pp. 1--17, 2021.

\bibitem{ding2022tdm}
J.~Ding, Z.~Wang, W.~Ma, X.~Wu, and M.~Wang, ``Tdm-mimo automotive radar
  point-cloud detection based on the 2-d hybrid sparse antenna array,''
  \emph{IEEE Transactions on Geoscience and Remote Sensing}, vol.~60, pp.
  1--15, 2022.

\bibitem{dvorsky2023multistatic}
M.~Dvorsky, S.~Y. Sim, D.~T. Motes, T.~Watt, A.~Shah, M.~T. Al~Qaseer, and
  R.~Zoughi, ``Multistatic ka-band (26.5--40 ghz) millimeter-wave 3-d imaging
  system,'' \emph{IEEE Transactions on Instrumentation and Measurement},
  vol.~72, pp. 1--14, 2023.

\bibitem{kong2024survey}
H.~Kong, C.~Huang, J.~Yu, and X.~Shen, ``A survey of mmwave radar-based sensing
  in autonomous vehicles, smart homes and industry,'' \emph{IEEE Communications
  Surveys \& Tutorials}, 2024.

\bibitem{park2024spatial}
J.-H. Park, S.~Lee, G.~Moon, and S.-C. Kim, ``Spatial-wideband effect
  compensation for high resolution imaging in mimo fmcw radar,'' \emph{IEEE
  Transactions on Instrumentation and Measurement}, 2024.

\bibitem{rai2025low}
C.~Rai and A.~Chattopadhyay, ``Low-complexity super-resolution signature
  estimation of xl-mimo fmcw radar,'' \emph{arXiv preprint arXiv:2506.07979},
  2025.

\bibitem{rabaste2013signal}
O.~Rabaste, L.~Savy, M.~Cattenoz, and J.-P. Guyvarch, ``Signal waveforms and
  range/angle coupling in coherent colocated mimo radar,'' in \emph{2013
  International Conference on Radar}.\hskip 1em plus 0.5em minus 0.4em\relax
  IEEE, 2013, pp. 157--162.

\bibitem{durr2020range}
A.~D{\"u}rr, B.~Schneele, D.~Schwarz, and C.~Waldschmidt, ``Range-angle
  coupling and near-field effects of very large arrays in mm-wave imaging
  radars,'' \emph{IEEE Transactions on Microwave Theory and Techniques},
  vol.~69, no.~1, pp. 262--270, 2020.

\bibitem{han2023range}
K.~Han and S.~Hong, ``Range-angle decoupling technique using
  wavelength-dependent beamforming for high-resolution mimo radar,'' \emph{IEEE
  Transactions on Microwave Theory and Techniques}, 2023.

\bibitem{hu2023range}
Y.~Hu, W.~Deng, Y.~Dong, and X.~Wu, ``Range-angle coupling in linear sparse
  array: Far-field model with narrow-band fmcw signal,'' \emph{IEEE
  Transactions on Aerospace and Electronic Systems}, 2023.

\bibitem{cai2017beamforming}
M.~Cai, J.~N. Laneman, and B.~Hochwald, ``Beamforming codebook compensation for
  beam squint with channel capacity constraint,'' in \emph{2017 IEEE
  International Symposium on Information Theory (ISIT)}.\hskip 1em plus 0.5em
  minus 0.4em\relax IEEE, 2017, pp. 76--80.

\bibitem{wang2018spatial}
B.~Wang, F.~Gao, S.~Jin, H.~Lin, and G.~Y. Li, ``Spatial-and frequency-wideband
  effects in millimeter-wave massive mimo systems,'' \emph{IEEE Transactions on
  Signal Processing}, vol.~66, no.~13, pp. 3393--3406, 2018.

\bibitem{wang2019beam}
B.~Wang, M.~Jian, F.~Gao, G.~Y. Li, and H.~Lin, ``Beam squint and channel
  estimation for wideband mmwave massive mimo-ofdm systems,'' \emph{IEEE
  transactions on signal processing}, vol.~67, no.~23, pp. 5893--5908, 2019.

\bibitem{rai2022signature}
C.~Rai and D.~Sen, ``Signature estimation of dual wideband systems,'' in
  \emph{2022 IEEE 95th Vehicular Technology Conference:(VTC2022-Spring)}.\hskip
  1em plus 0.5em minus 0.4em\relax IEEE, 2022, pp. 1--5.

\bibitem{rai2025multi}
C.~Rai, H.~Singh, and A.~Chattopadhyay, ``Multi-target range, doppler and angle
  estimation in mimo-fmcw radar with limited measurements,'' \emph{arXiv
  preprint arXiv:2502.01147}, 2025.

\bibitem{rai2025two}
C.~Rai and D.~Sen, ``A two-stage rotation-based super-resolution signature
  estimation for spatial wideband systems,'' \emph{arXiv preprint
  arXiv:2503.18111}, 2025.

\bibitem{rai2026low}
C.~Rai and A.~Chattopadhyay, ``Low-complexity super-resolution signature
  estimation of xl-mimo fmcw radar,'' \emph{IEEE Transactions on Aerospace and
  Electronic Systems}, 2026.

\bibitem{wang2006novel}
J.~Wang and Z.~Shen, ``A novel method for two-dimensional frequency
  estimation,'' \emph{IEEE Transactions on Circuits and Systems II: Express
  Briefs}, vol.~53, no.~2, pp. 148--151, 2006.

\bibitem{bhaskar2013atomic}
B.~N. Bhaskar, G.~Tang, and B.~Recht, ``Atomic norm denoising with applications
  to line spectral estimation,'' \emph{IEEE Transactions on Signal Processing},
  vol.~61, no.~23, pp. 5987--5999, 2013.

\bibitem{chi2014compressive}
Y.~Chi and Y.~Chen, ``Compressive two-dimensional harmonic retrieval via atomic
  norm minimization,'' \emph{IEEE Transactions on Signal Processing}, vol.~63,
  no.~4, pp. 1030--1042, 2014.

\bibitem{rai2025low1}
C.~Rai and D.~Sen, ``Low complexity doa-toa signature estimation for
  multi-antenna multi-carrier systems,'' in \emph{ICASSP 2025-2025 IEEE
  International Conference on Acoustics, Speech and Signal Processing
  (ICASSP)}.\hskip 1em plus 0.5em minus 0.4em\relax IEEE, 2025, pp. 1--5.

\bibitem{zhang2019efficient}
Z.~Zhang, Y.~Wang, and Z.~Tian, ``Efficient two-dimensional line spectrum
  estimation based on decoupled atomic norm minimization,'' \emph{Signal
  Processing}, vol. 163, pp. 95--106, 2019.

\bibitem{zhang2017decoupled}
Z.~Zhang, W.~Wang, Y.~Huang, and S.~Liu, ``Decoupled 2-d direction of arrival
  estimation in l-shaped array,'' \emph{IEEE Communications Letters}, vol.~21,
  no.~9, pp. 1989--1992, 2017.

\bibitem{hu2022decoupled}
Y.~Hu, P.~N. Samarasinghe, S.~Gannot, and T.~D. Abhayapala, ``Decoupled
  multiple speaker direction-of-arrival estimator under reverberant
  environments,'' \emph{IEEE/ACM Transactions on Audio, Speech, and Language
  Processing}, vol.~30, pp. 3120--3133, 2022.

\bibitem{weng2023wideband}
S.~Weng, F.~Jiang, and H.~Wymeersch, ``Wideband mmwave massive mimo channel
  estimation and localization,'' \emph{IEEE Wireless Communications Letters},
  vol.~12, no.~8, pp. 1314--1318, 2023.

\bibitem{rai2023sparse}
C.~Rai and D.~Sen, ``Sparse scatter/target detection with spatial wideband
  uniform linear arrays,'' in \emph{2023 IEEE 97th Vehicular Technology
  Conference (VTC2023-Spring)}.\hskip 1em plus 0.5em minus 0.4em\relax IEEE,
  2023, pp. 1--5.

\bibitem{lovescu2020fundamentals}
C.~Lovescu and S.~Rao, ``The fundamentals of millimeter wave radar sensors,''
  \emph{Texas Instruments, Julio}, 2020.

\bibitem{wang2017clustering}
Z.~Wang, Z.~Yu, C.~P. Chen, J.~You, T.~Gu, H.-S. Wong, and J.~Zhang,
  ``Clustering by local gravitation,'' \emph{IEEE transactions on cybernetics},
  vol.~48, no.~5, pp. 1383--1396, 2017.

\end{thebibliography}

\end{document}